# Imaging the Transverse Component of Optical Near-Fields in Resonant Photonic Structures

Petr Koutenský,* Neli Laštovičková Streshkova, Stefanie Kraus, Peter Hommelhoff, and Martin Kozák*

**ABSTRACT:** We report on imaging the optical near-fields in resonant periodic photonic structures with nanometer resolution using ultrafast 4D scanning transmission electron microscopy (U4DSTEM). In particular, U4DSTEM is applied to visualize the transverse component of the Lorentz force of a synchronous near-field mode excited by an infrared femtosecond pulse in a periodic silicon nanostructure designed for the photonic acceleration of electrons. Our results show that in addition to the accelerating/decelerating force acting on the electrons in the longitudinal direction along the electron propagation, the structures can be efficiently used for transverse electron streaking at optical frequencies when excited by light with polarization perpendicular to the electron trajectory. The measured spatial profile of the excited near-field mode intensity is consistent with the numerical simulations performed using the finite-difference time-domain technique.

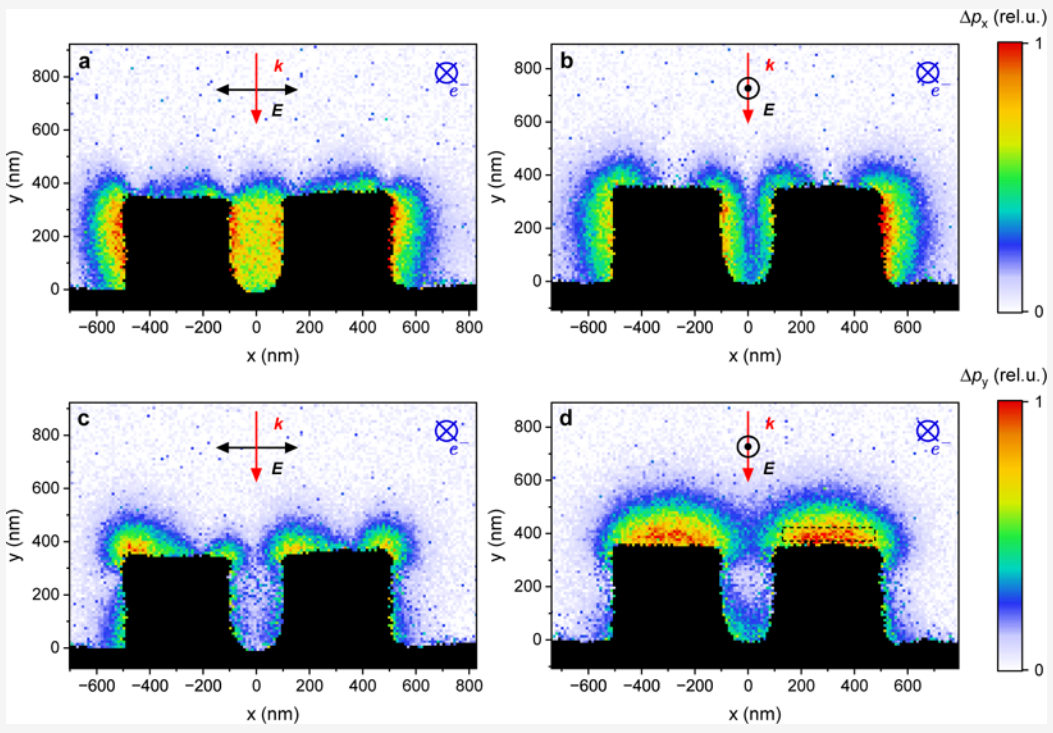

## 1. INTRODUCTION

Photonics utilizes nanostructures with feature sizes smaller than the wavelength of light to manipulate the functionality of the materials by structuring the local electromagnetic fields.[1,2] Such manipulation allows one to modify the spectral response of a material and control its linear and nonlinear optical properties without the need to change the composition or atomic structure of the material.[2−4] Closely related to the field of photonics is plasmonics, focusing on the collective electromagnetic oscillations of electrons and the electromagnetic field of light.[5] Photonic and plasmonic metamaterials play an indispensable role in many fields of science, including optics, biophysics, and chemistry.

All these fields have one joint property, the subwavelength localization of the electromagnetic near-field. To understand the functionality of photonic and plasmonic structures, it is essential to utilize experimental and theoretical tools capable of visualizing near-field distributions with deep subwavelength resolution. Standard optical microscopy is insufficient for this task because of the diffraction-limited spatial resolution and because it is sensitive only to the spatial variation of the dielectric function but not to the optical near-field itself. Although near-field scanning optical microscopy can deliver information about optical fields at surfaces with subwavelength precision,[6] it fails to image optical or plasmonic fields deep within the layers of photonic structures. In contrast, transmission-based electron microscopy is well adapted for such a task, as the electrons can propagate through the hollow parts of the structures, providing a direct probe of the internal electromagnetic environment.

There have been several methods utilizing electron beams that are capable of visualizing optical or plasmonic near-fields. From the point of view of electron optics, the simplest method is based on point-projection microscopy using electrons photoemitted from a nanotip and accelerated by the applied static field.[7] However, such imaging has severe limitations and does not allow one to resolve the intensity distribution of the near-field. This goal has been reached by photon-induced near-field electron microscopy (PINEM), which has been developed to visualize the optical near-fields generated at the surfaces of nanostructures illuminated by coherent light in transmission electron microscopes. PINEM is based on energy-resolved imaging of inelastically scattered electrons that exchanged energy with localized light modes in units of light quanta.[8−28] PINEM can be experimentally realized either with a focused electron beam or with a collimated electron beam. In the first case, scanning transmission electron microscopy is combined with electron energy loss spectroscopy (STEM-EELS), and an EELS spectrum is recorded for each point scanned in the sample plane. The second method, energy-filtered transmission electron microscopy (EFTEM), utilizes standard projection-based imaging combined with spectral filtering of the electrons. Because the component of the electron momentum change in PINEM is the one along the electron propagation direction, this method is sensitive to the longitudinal component of the Lorentz force. Moreover, due to the necessity to fulfill energy and momentum conservation during inelastic electron scattering,[8] only near-field modes whose phase velocity is synchronized with the group velocity of the electrons can efficiently modulate the electron momentum and can be visualized by PINEM. The amplitude of the longitudinal component of the Lorentz force, $F_z = eE_z$, which is proportional to the longitudinal near-field component $E_z$, is not constant, but it is inhomogeneous in the transverse direction. The transverse gradients of $E_z$ are directly related to the transverse electron deflections, as explained by the Panofsky−Wenzel theorem,[29] whose application to PINEM interaction is discussed in detail in the next chapter. However, the transverse deflection of electrons is only rarely characterized in PINEM-type experiments[17,21,28] due to the small angles of deflection in TEMs, where PINEM is most commonly implemented.

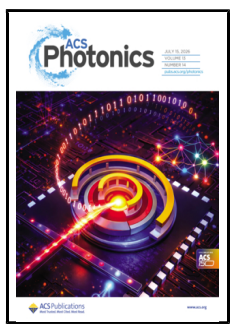

To characterize the transverse components of the Lorentz force acting on electrons when propagating inside photonic structures, one can use a modification of Lorentz force microscopy, which was developed to image static electric or magnetic fields in samples using a continuous electron beam.[30] Here, the electron deflection angle is measured as a function of the beam position on the sample, resulting in a map of the transverse Lorentz force integrated along the electron trajectory, which modifies the transverse momentum of the electrons. The method has recently been transferred to ultrafast electron microscopy[28,31] and formed the basis for ultrafast scanning transmission electron microscopy (U4DSTEM).[32−35] When coherent light that excites the near-field has the form of a femtosecond laser pulse, the amplitude of the electric field in the near-field region can reach values of the order of 1 V/nm. This field is sufficient to deflect the electrons by a few mrads even when the interaction takes place over a very short distance of a few tens of nanometers.[35] In U4DSTEM, the electron beam focused on the nanostructure is scanned through the near-field region, and the resulting two-dimensional electron scattering patterns are detected using a hybrid pixel detector as a function of the beam position in the sample plane. Based on the shape of the pattern formed by scattered electrons, the local field strength and direction are calculated for each position of the electron beam in the sample plane. PINEM is typically implemented in an ultrafast transmission electron microscope due to the requirement of energy-filtered imaging, and only a few works show its application in the low electron energy regime.[27] In contrast, U4DSTEM has recently been demonstrated in a modified scanning electron microscope,[35] which represents a more accessible tool, which is routinely available in most laboratories in the fields of material science, photonics, and nanofabrication. Both PINEM and U4DSTEM offer deeply subwavelength spatial resolution and temporal resolution in the femtosecond regime.

In this work, we apply the U4DSTEM technique to image the transverse components of the Lorentz force generated by optical near-fields in periodic photonic structures made of crystalline silicon. Such structures were designed for electron acceleration by light utilizing the inverse Smith-Purcell effect.[36−43] The principle of efficient acceleration is based on the synchronization of the phase velocity of a particular spatial harmonic component of the near-field with the propagating electron. Due to the extended interaction distance between the electron and light compared to a single nanostructure without periodicity, the probability of stimulated absorption and emission of photons by the electron is strongly enhanced.[37,44] The resonant interaction between the electron and light allows the electron to absorb or emit the energy corresponding up to $\approx 10^4$ photons. As a result of transverse forces of the synchronous mode, the electrons are deflected to angles of an order of magnitude greater than in similar previous studies,[45,46] greatly improving the sensitivity of the method. For these reasons, quantization of electron energy modulation does not play a significant role, and the interaction can typically be described fully classically.[36−38,42,43]

The optical near-fields in the vicinity of complex-shaped nanostructures can be calculated by numerical solution of Maxwell's equations using finite-difference time-domain (FDTD) or by other numerical methods. However, the numerical simulations are usually idealized and do not account for, e.g., real-world structural deviations introduced during nanofabrication, deviations caused by changes of the dielectric function of the nanostructured material in surface layers with different morphology, or bulk contamination during production. Characterization of longitudinal fields in resonant periodic photonic nanostructures using PINEM[47] showed that such effects lead to small distortions in the spatial profile of the field compared to numerical simulations. However, the transverse field components that determine the electron trajectories and beam stability during acceleration inside a narrow channel remain largely unexplored.[48] U4DSTEM provides information on the missing components of the electromagnetic fields and bridges the gap between idealized simulation and fabricated reality by providing a direct, high-resolution map of the forces an electron actually experiences within the device. Combined with PINEM, this method will allow characterizing all three components of the Lorentz force, enabling a full 3D reconstruction of the optical near-field in the future.

## 2. THEORY OF TRANSVERSE ELECTRON DEFLECTION IN OPTICAL NEAR-FIELDS

The interaction between an electron and electromagnetic fields in a vacuum can be described classically, semiclassically, or by a fully quantum approach. In PINEM, the electron wave function after the interaction with light is represented by an equidistant ladder of states in the energy domain, which results from the absorption and emission of photon quanta from the light field.[8,9,11] In the semiclassical picture and within the nonrecoil approximation, the electron spectrum with separated photon sidebands originates from a periodic phase modulation of the electron wave function. The wave function after the interaction with a localized optical near-field at frequency $\omega$ can be written as

$$\phi(z', t) = \phi_0(z', t_0)\exp[i2|\beta|\sin(\omega z'/v + \arg(\beta))] \quad (1)$$

where the coupling constant $\beta$ is given by

$$\beta(\omega/v) = \frac{e}{2\hbar\omega}\int_{-\infty}^{\infty}\tilde{E}_z(z)\exp[-i\omega z/v]\mathrm{d}z \quad (2)$$

The electron propagating along the $z$ direction at velocity $v$ effectively interacts only with the longitudinal electric field component $E_z$ of a mode with phase velocity matched to the electron group velocity. The contributions of all other spatial components cancel out after integration in eq 2, which represents a Fourier transform between real and reciprocal spaces. Because the coupling occurs strictly between the electron phase and the $E_z$ component of the field, the origin of the transverse deflection of the electron originating in the semiclassical picture is not obvious. It can be understood in the framework of the Panofsky–Wenzel theorem,[29] which relates the transverse components of the electron momentum change with the transverse derivatives of the longitudinal force components acting on the electron. While it was first derived for classical relativistic particles, we show in the following that it can be applied to the case of the PINEM interaction.

The transverse and longitudinal components of the classical Lorentz force are not independent as a consequence of the Faraday's law:

$$-\frac{\partial \mathbf{B}}{\partial t} = \nabla \times \mathbf{E} \tag{3}$$

Writing the explicit equations for the first two vector components of the curl gives

$$\begin{aligned} \frac{\partial E_y}{\partial z} + v\frac{\partial B_x}{\partial z} - \frac{\mathrm{d}B_x}{\mathrm{d}t} &= \frac{\partial E_z}{\partial y}, \\ \frac{\partial E_x}{\partial z} - v\frac{\partial B_y}{\partial z} + \frac{\mathrm{d}B_y}{\mathrm{d}t} &= \frac{\partial E_z}{\partial x}, \end{aligned} \tag{4}$$

where we assume that the electron travels along the $z$ axis $\mathbf{v} = (0, 0, v)$ and we express the partial derivative in $t$ along the electron trajectory $z = vt$ in terms of the total differential $\mathrm{d}\mathbf{B}_i/\mathrm{d}t = \partial \mathbf{B}_i/\partial t + v\partial \mathbf{B}_i/\partial z$. After multiplication by $e$, the first two terms on each line correspond to a transverse component of the Lorentz force, and eq 4 can be rewritten in a vector form:

$$\frac{\partial \mathbf{F}_\perp}{\partial z} - e\left(\hat{z} \times \frac{\mathrm{d}\mathbf{B}}{\mathrm{d}t}\right)_\perp = e\nabla_\perp E_z \tag{5}$$

To obtain the total momentum change of the electron, we integrate both sides of eq 5 over the interaction time, respectively, over the electron trajectory $\mathrm{d}t = \mathrm{d}z/v$. The magnetic field vanishes far away from the near-field, $\mathbf{B}[\infty] = \mathbf{B}[-\infty] = 0$, and so does the contribution of $\mathrm{d}\mathbf{B}/\mathrm{d}t$ under the integral. The rest yields the Panofsky–Wenzel theorem[29] for PINEM-type interaction:

$$\frac{\partial}{\partial z}\Delta \mathbf{p}_\perp = \frac{1}{v}\int_{-\infty}^{\infty} \frac{\partial \mathbf{F}_\perp}{\partial z}\mathrm{d}z = \frac{e}{v}\int_{-\infty}^{\infty} \nabla_\perp E_z \mathrm{d}z = \nabla_\perp \Delta p_z \tag{6}$$

The $z$ gradient of the transverse momentum change $\Delta \boldsymbol{p}_\perp$ is connected to the transverse gradients $\nabla_\perp E_z$ integrated along the electron trajectory, which we recognize as the longitudinal momentum change $\Delta p_z$. The same relation holds for semiclassical description, where the classical momentum is replaced by the expectation value of the momentum operator. In the semiclassical picture, the near-field imprints a spatially inhomogeneous phase profile on the electron wave function proportional to $E_z$,[8] and its transverse gradients $\nabla_\perp E_z$ lead to transverse deflections. However, these interpretations are equivalent. We note that the Panofsky–Wenzel theorem applied to the PINEM case does not express the local relations between the longitudinal and transverse components of the electric and magnetic fields but only holds for the induced momentum change.

Another interesting question is whether we may expect that the transverse part of the wave function of the electron after the interaction with the near-field can be represented by a superposition of discrete transverse momentum states similar to the case of the longitudinal momentum in classical PINEM. The harmonic phase modulation in the longitudinal direction can be attributed to the harmonic oscillations of the optical near-fields in the time domain, which generate a spatially periodic potential in the rest frame of a moving electron, with its periodicity aligned with the electron propagation direction. In contrast, in the case of a general near-field distribution, there is no periodicity in the transverse directions (perpendicular to electron beam propagation), which could lead to coherent diffraction peaks in the electron scattering pattern. The only exception is the case where the field has its own periodicity with a period smaller than the transverse coherence length of the electron beam.[21] However, the electron beam is typically focused in both PINEM (STEM-EELS) and U4DSTEM to spatial dimensions much smaller than the wavelength of the excitation light, preventing the possible coherent interference of electron waves coming from different periods of the near-field. For this reason, the electron wave is sensitive only to a local amplitude and gradient of the electromagnetic potential, but no photon peaks are expected in the transverse-momentum distribution of scattered electrons. In this regime, the quantum mechanical and classical descriptions of the interaction lead to the same transverse momentum distributions of electrons, and the interaction between the electron and optical near-field can be described classically in a point-particle approximation.[35]

In the classical picture, the transverse deflection of the electron can be explained as the effect of the transverse components of the Lorentz force $\mathbf{F}_\perp$ acting on the point-particle electron during its propagation through the near-fields. In the electron rest frame, the $j$ component of the total change in the electron transverse momentum in spatial coordinates $x$, $y$ is

$$\Delta p_j(x, y) = \int_{-\infty}^{\infty} F_j(x, y, t)\mathrm{d}t = N\int_T F_j(x, y, t)\mathrm{d}t \tag{7}$$

where $F_j = e[\mathbf{E} + (\mathbf{v}\times\mathbf{B})]_j$ denotes the $j$th component of the Lorentz force with the electric and magnetic fields $\mathbf{E} = \Re\{\tilde{\mathbf{E}}(\mathbf{r}, \omega)g(t - \Delta t)e^{i\omega t+i\varphi_0}\}$ and $\mathbf{B} = \Re\{\tilde{\mathbf{B}}(\mathbf{r}, \omega)g(t - \Delta t)e^{i\omega t+i\varphi_0}\}$, respectively. Here, $\Delta t$ represents the electron arrival time with respect to the optical field, and $g(t - \Delta t)$ and $\varphi_0$ are the envelope and phase of the optical fields, respectively. The spatial distributions of the electromagnetic fields $\tilde{\mathbf{E}}(\mathbf{r},\omega)$ and $\tilde{\mathbf{B}}(\mathbf{r},\omega)$ at angular frequency $\omega$ are obtained from a numerical solution of Maxwell's equations (for details, see Supporting Information, section Numerical simulations of electromagnetic field distribution). In the case of synchronous interaction of an electron with a near-field mode of a resonant periodic photonic structure investigated in this work, the integration over the electron trajectory can be replaced by integration over a single period of length $\Lambda = vT$, where $v$ is the electron velocity and $T$ is the time period of the driving coherent light, multiplied by the number of periods $N$. Here, we assume a nonrecoil approximation (electron trajectory does not change signifi-cantly during the interaction), and we neglect edge effects caused by the finite length of the periodic structure.

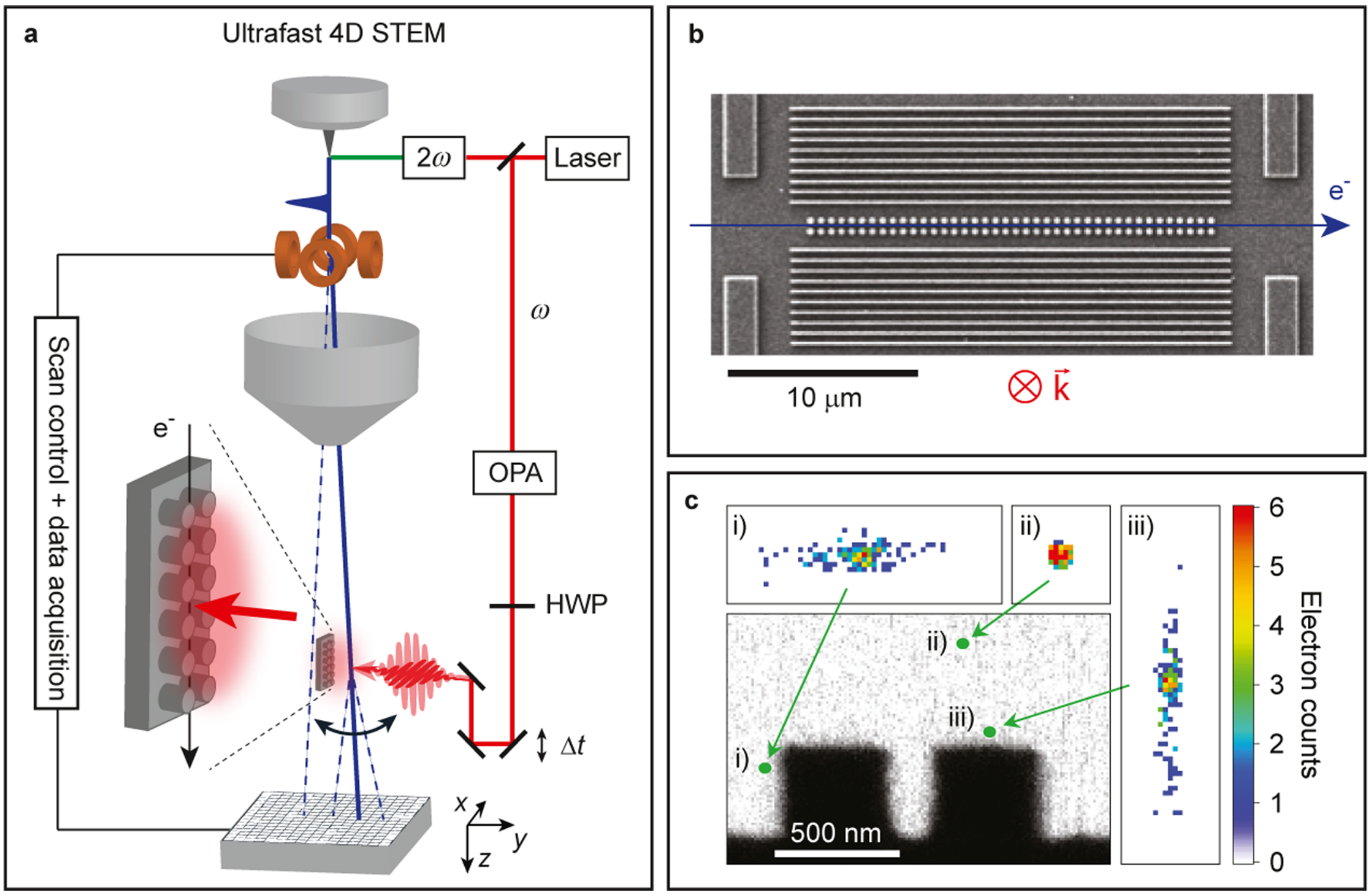

**Figure 1.** Ultrafast 4D scanning transmission electron microscopy (U4DSTEM). (a) Layout of the U4DSTEM experimental setup. Second harmonics generated by a fraction of the fundamental laser output generate electron pulses in the electron gun of a scanning electron microscope. An electron beam is focused and scanned across a periodic nanostructure. Electrons transmitted in the vicinity of the structure surface interact with the synchronous mode of the optical near-field generated by an infrared pulse generated in an optical parametric amplifier (OPA) with the wavelength of 1.93 $\mu$m. Excitation light has linear polarization with the direction adjusted by a half-wave plate (HWP) to be parallel or perpendicular to the electron propagation direction. Electron deflection/transverse scattering is characterized using a hybrid-pixel detector placed 170 mm downstream from the nanostructure. (b) Scanning electron microscopy (SEM) image of the periodic nanostructure. The blue arrow indicates the trajectory of electrons. The red cross illustrates the direction of the pointing vector of the excitation infrared light pulse. (c) Scanning transmission electron microscopy (STEM) image of the nanostructure (dark shadow). Electron scattering patterns measured in three different electron beam positions labeled by the green dots are shown in the insets.

## 3. RESULTS

The layout of the experimental setup is shown in Figure 1a. The resonant nanostructure placed in the focal plane of an ultrafast scanning electron microscope has its periodicity parallel to the trajectory of the electrons. A pulsed laser beam with a wavelength of 1930 nm and a pulse duration of ≈110 fs incident from the direction perpendicular to the substrate excites the optical near-field. The excitation light is linearly polarized along either the direction of propagation of electrons or perpendicular to it. The pulsed electron beam with a kinetic energy of 28.6 keV, a repetition rate of 500 kHz, and a pulse duration of 800 fs at the sample is focused on the nanostructure located at a working distance of the electron microscope of 15 mm. The electron acceleration voltage is chosen such that the electrons interact in phase with the optical field in every period of the resonant nanostructure. To generate an almost collimated electron beam with a current sufficient for U4DSTEM imaging, we use the highest probe current setting of the electron microscope. The beam divergence is reduced by introducing an objective lens aperture with a diameter of 64 $\mu$m. The resulting electron beam has a spot size in the focus of ≈21 nm and a semidivergence angle of 1 mrad.

The photonic nanostructure under study (Figure 1b) was prepared using electron lithography and reactive ion etching. It consists of 2 columns of cylinders with a diameter of 400 nm periodically spaced with a period of 620 nm and a height of 350 nm. The gap between the two columns of the pillars was 225 nm. To enhance the coupling between the incident field and the synchronous near-field mode, the dual-pillar structure is surrounded on both sides by Bragg mirrors consisting of 10 etched trenches. The width of each trench is 225 nm, and their period is 708 nm. The distance between the edge of the first step and the edge of the closest cylinder column was 848 nm. More information related to this photonic nanostructure and its fabrication can be found in ref 49.

The transverse momentum change of the electrons corresponding to the strength of the transverse component of the Lorentz force integrated over the duration of the interaction with the near-field is measured as a function of the position of the electron beam in the sample plane by detecting the scattered electron images (examples are shown in Figure 1c) using a hybrid pixel detector TimePix3. The electron deflection angle is directly proportional to the transverse momentum change. There is a broad distribution of the measured deflection angles owing to three factors: (i) the electron pulse duration is longer than the envelope of the driving laser pulse, (ii) the electrons interact with a random phase of the optical near-field, and (iii) the incident electron beam has a finite semidivergence angle. To extract the transverse component of the change in the electron momentum $\Delta p_j$ from the measured data, we calculate the standard deviation of the transverse momentum distribution of the detected electrons in each pixel $\sigma_{\Delta p_j}$ and use the relation between the standard deviation and the average change in electron momentum $\Delta p_j^2 = \sigma_{\Delta p_j}^2 - p_e^2/4$, where $p_e$ is the radius of the initial electron density in the transverse momentum space, representing the angular divergence of the electron beam (see Supporting Information, Analytical formula variance).

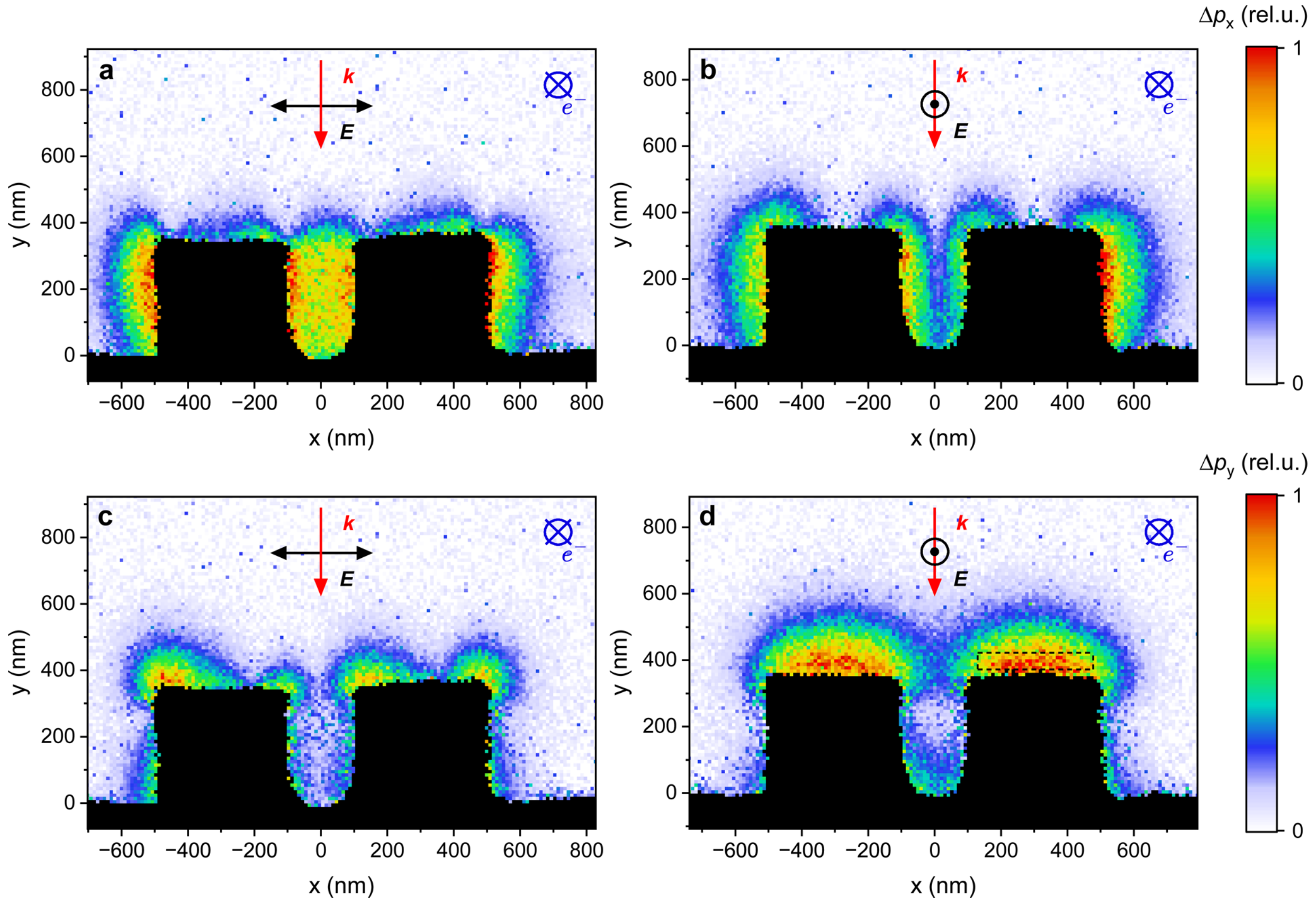

**Figure 2.** Images of the transverse component of Lorentz force induced by an optical near-field generated on a surface of a periodic photonic structure. (a,c) Linear polarization of excitation light is perpendicular to the electron propagation. (b,d) Linear polarization is parallel to the electron trajectory. The red arrows indicate the wave vector of coherent optical excitation, the black arrows depict its polarization direction, and the blue crosses represent the direction of electron beam propagation (into the screen).

Figure 2 shows the measured images of the transverse components of the electron momentum change $\Delta p_x$ and $\Delta p_y$ induced by the optical near-field for two different orientations of linear polarization of excitation light. The red arrow denotes the wave vector of light $\vec{k}$, the black arrow its polarization $\vec{E}$, and the blue-arrow crosses represent the direction of propagation of the electrons.

For comparison, Figure 3 shows the transverse components of the electron momentum change calculated using the spatiotemporal distribution of electromagnetic fields obtained by finite-difference time-domain simulations using the commercial software Lumerical FDTD (see Supporting Information, Numerical simulations). In Figures 2a,c and 3a,c, the excitation light is polarized perpendicular to the trajectory of the electrons, while in Figures 2b,d and 3b,d, the polarization points along the electron trajectory. We see a qualitatively different behavior of the transverse force components of the near-field for the two different polarization directions. In the case of the electric field polarization pointing in the direction of the electrons, the force acting on the electrons inside the channel is along the initial field polarization, and deflection of electrons in the center of the channel between the pillars is zero. In this case, the electrons are mainly deflected when propagating along the top surface of the pillars or along the outer edges of the pillars. In contrast, when the polarization of the excitation field is rotated by 90°, the electrons are strongly deflected by the optical near-field in the center of the channel in the direction of incident field polarization.

The photonic structures used in this work were designed primarily for efficient acceleration of electrons.[49] However, our U4DSTEM results show that these structures can work for both efficient energy modulation and transverse streaking of subrelativistic electrons at optical frequencies, allowing the application of this type of photonic structure for attosecond experiments with subrelativistic electrons.[40,46,50−53]

We note that there is one important difference between PINEM and U4DSTEM imaging of the near-fields. In the case of inelastic electron scattering detected in PINEM, the longitudinal momentum of the electrons is modulated only by the interaction with the electric field component along its propagation. The reason is that when an electron propagates with $\mathbf{v}$, the projection of the magnetic part of the Lorentz force to the direction of propagation $(\mathbf{v}/v).(\mathbf{v} \times \mathbf{B})$ is zero. The magnetic part of the force thus cannot accelerate or decelerate the electrons, regardless of the direction of the magnetic field. In contrast, the electron deflection corresponding to the transverse momentum change, which is measured in U4DSTEM, is sensitive to both parts of the Lorentz force. This is the reason why the measured transverse momentum change cannot be straightforwardly related only to the electric field distribution but rather shows the distribution of the total Lorentz force acting on the electrons.

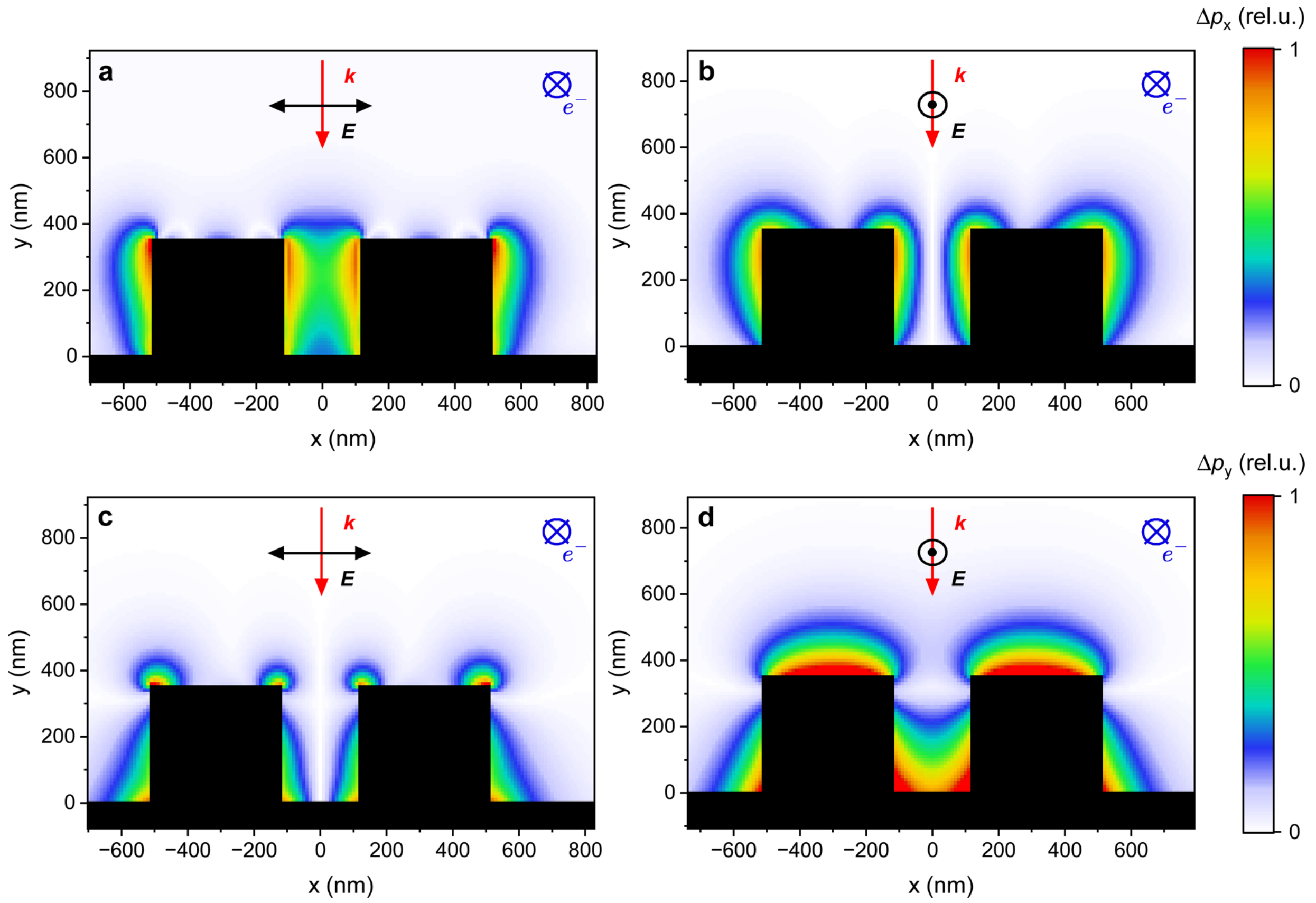

**Figure 3.** Simulations of the transverse component of Lorentz force induced by an optical near-field generated on a surface of a periodic photonic structure. (a,c) Linear polarization of excitation light is perpendicular to the electron propagation. (b,d) Linear polarization is parallel to the electron trajectory. The red arrows indicate the wave vector of coherent optical excitation, the black arrows depict its polarization direction, and the blue crosses represent the direction of electron beam propagation (into the screen).

Near the surface of the substrate, the transverse momentum change measured experimentally is noticeably smaller than the one obtained from the simulations for both polarizations. This is most likely caused by collisions of part of the deflected electron distribution with the substrate, which is 100 $\mu$m long. When the electron beam position is 50 nm from the substrate, the electrons deflected by more than ≈1 mrad collide with the substrate and are not detected, leading to an artificially smaller standard deviation of the measured electron distribution.

The coupling between electrons and the optical near-field is linear, and the transverse momentum change should scale linearly with the amplitude of the electric field of the excitation light. Figure 4 shows the measured transverse momentum change $\Delta p_y$ as a function of the amplitude of the electric field of excitation $E$, which was polarized parallel to the direction of electron propagation. The integrated $E_y$ was obtained as an arithmetic mean from the region marked by a dashed black rectangle in Figure 2d. Measurements were conducted for excitation powers of 10, 20, 30, 50, and 120 mW. Values of $E$ and $E_y$ are in relative units and normalized with respect to the maximum.

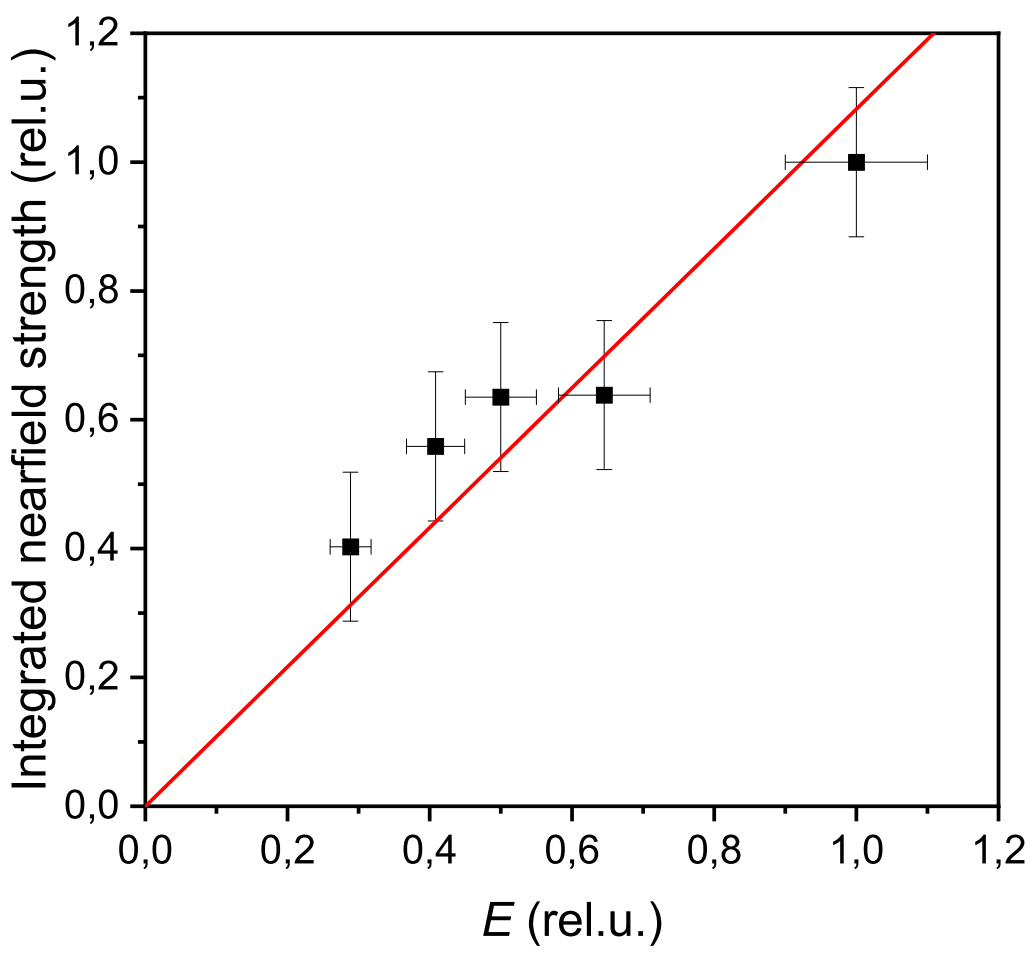

**Figure 4.** Integrated nearfield Lorentz force as a function of the amplitude of the excitation field. The excitation light was polarized parallel to the trajectory of electrons. The measurements were conducted for excitation powers of 10, 20, 30, 50, and 120 mW. The plotted integrated strength is obtained as an arithmetic mean from the dashed region in Figure 2d. All values are normalized.

## 4. DISCUSSION

We demonstrate that U4DSTEM is a powerful imaging technique, which can be applied to visualize transverse components of optical near-fields of resonant nanostructures. Due to the complementary information with respect to PINEM, where the longitudinal near-field component is characterized, a combination of these two methods paves the way toward complete reconstruction of all components of optical and plasmonic near-fields with deep subwavelength resolution. Our results show that the transverse streaking at optical frequencies can be strongly enhanced by using periodic dielectric structures, enabling a significant improvement in time resolution in future attosecond experiments with free electrons.

## ■ ASSOCIATED CONTENT

### Supporting Information

The Supporting Information is available free of charge at https://pubs.acs.org/doi/10.1021/acsphotonics.6c00984.

Additional details of the experimental setup, analytical formula for transverse momentum change, and numerical simulations of electromagnetic field distribution (PDF)

## ■ AUTHOR INFORMATION

### Corresponding Authors

**Petr Koutenský** − *Department of Chemical Physics and Optics, Faculty of Mathematics and Physics, Charles University, Prague 2 CZ-12116, Czech Republic;* orcid.org/0000-0003-4851-2331; Email: petr.koutensky@matfyz.cuni.cz

**Martin Kozák** − *Department of Chemical Physics and Optics, Faculty of Mathematics and Physics, Charles University, Prague 2 CZ-12116, Czech Republic;* orcid.org/0000-0002-6317-7079; Email: m.kozak@matfyz.cuni.cz

### Authors

**Neli Laštovičková Streshkova** − *Department of Chemical Physics and Optics, Faculty of Mathematics and Physics, Charles University, Prague 2 CZ-12116, Czech Republic;* orcid.org/0009-0004-3931-3597

**Stefanie Kraus** − *Physics Department, Friedrich Alexander University Erlangen-Nürnberg, Erlangen 91058, Germany*

**Peter Hommelhoff** − *Physics Department, Friedrich Alexander University Erlangen-Nürnberg, Erlangen 91058, Germany; Faculty of Physics, Ludwig Maximilian University Munich, Munich 80799, Germany*

Complete contact information is available at: https://pubs.acs.org/10.1021/acsphotonics.6c00984

### Funding

The authors acknowledge funding from the Czech Science Foundation (project 22-13001K), Charles University (SVV-2023-260720, PRIMUS/19/SCI/05, GAUK 90424), the European Union (ERC, eWaveShaper, 101039339, and ERC, AccelOnChip, 884217), and DFG (HO 4543/5, HO 4543/7, and HO 4543/8). This work was supported by the TERAFIT project No. CZ.02.01.01/00/22_008/0004594 funded by OP JAK, Excellent Research call.

### Notes

The authors declare no competing financial interest.
The preprint version of this work can be found on arXiv.[55]

## ■ ACKNOWLEDGMENTS

The authors acknowledge the contribution of P. Yousefi, who fabricated the nanostructures used in this study. All of the data that support the plots and the other findings of this study are publicly available at DOI: 10.5281/zenodo.19662201.[54]

## ■ REFERENCES